# Aluminum nitride waveguide beam splitters for integrated quantum photonic circuits

**HYEONG-SOON JANG,**[1,2] **DONGHWA LEE,**[1,3] **, HYUNGJUN HEO,**[1] **YONG-SU KIM,**[1,3] **HYANG-TAG LIM,**[1,3] **SEUNG-WOO JEON,**[1] **SUNG MOON,**[1,3] **SANGIN KIM,**[2] **SANG-WOOK HAN,**[1,3] **HOJOONG JUNG,**[1,2,*]

[1]*Center for Quantum Information, Korea Institute of Science and Technology (KIST), Seoul 02792, South Korea*
[2]*Department of Electrical and Computer Engineering, Ajou University, Suwon 16499, South Korea*
[3]*Division of Nano and Information Technology, KIST School, Korea University of Science and Technology, Seoul 02792, South Korea*
*\* hojoong.jung@kist.re.kr*

**Abstract:** We demonstrate integrated photonic circuits for quantum devices using sputtered polycrystalline aluminum nitride (AlN) on insulator. The on-chip AlN waveguide directional couplers, which are one of the most important components in quantum photonics, are fabricated and show the output power splitting ratios from 50:50 to 99:1. The polarization beam splitters with an extinction ratio of more than 10 dB are also realized from the AlN directional couplers. Using the fabricated AlN waveguide beam splitters, we observe the Hong-Ou-Mandel interference with a visibility of 91.7 ± 5.66 %.

## 1. Introduction

Recently, quantum supremacy has been demonstrated and triggered a lot of interest in the quantum information community [1-3]. Among many candidates for quantum processors, the photonic qubit based integrated quantum circuits are emerging as a promising platform for quantum information processing by replacing the bulk quantum optics system which already have shown its potential for a quantum computer and quantum simulator [4-7].

Traditionally, silicon has been widely used for integrated photonic circuits thanks to the mature nano-fabrication technology. For quantum device applications, however, low optical loss in a broad wavelength range and fast modulation speed are highly preferred where silicon has limited properties [8-10]. Other materials such as silica [11], silicon nitride (SiN) [12], Diamond [13], tantalum pentoxide ($Ta_2O_5$) [14] have limited modulation speed, and the lithium niobate ($LiNbO_3$) needs to be developed more for mass production [15]. Meanwhile, the group III-V materials including GaN, AlGaAs, InP and AlN provide fast electro-optic modulations and low enough optical losses [16-18]. Especially AlN shows the lowest optical loss among the group III-V materials and broadband transparent window from ultraviolet (200 nm) to mid-infrared (13.6 μm) [19-21]. These great optical property and electro-optic effect of AlN can be applied for integrated quantum photonics.

In this paper, we report a full set of AlN directional couplers that cover all beam splitting ratio, which is a key component for quantum photonic devices. We also fabricate polarization beam splitters by adjusting the coupling conditions of TE and TM modes in the directional couplers. Furthermore, a two-photon interference experiment is performed using the 50:50 directional coupler, and the Hong-Ou-Mandel (HOM) interference with 91.7 ± 5.66 % visibility is observed without accidental coincidence counts subtraction. Detailed fabrication methods and measurement process are explained in the following.

## 2. Single mode waveguide

The 500 nm-thick poly-crystalline AlN thin film is deposited on 3 μm thermally grown $SiO_2$ on 4 inch Si substrate by RF sputtering. Fig. 1(a) shows the measured refractive indices of the sputtered AlN thin film from 250 nm to 1650 nm wavelength range using the ellipsometry. The dashed lines are the indices of the as-deposited AlN, and the solid lines are the indices of AlN after annealing process at 950 °C in $N_2$ environment for 1 hour. After the annealing, both ordinary ($n_o$) and extraordinary ($n_e$) refractive indices are decreased over the whole measured wavelength range. The imaginary part of refractive index or extinction coefficient ($k$) of AlN at telecom band, which has to be small enough to be used for the quantum integrated circuit, is less than the reliable equipment measurement limit ($1\times10^{-6}$) even before annealing. The high extinction coefficient at ultraviolet regime (up to 0.01 or $3.5\times10^3$ dB/cm) decreases dramatically after annealing which indicates our annealing process reduces the optical absorption of the AlN thin film. We also qualitatively observe the decreased extinction coefficient from visible to telecom band after annealing even though it is too small to be measured accurately

To avoid modal dispersion, we engineer the geometry of AlN waveguides to support only the first order transverse electric (TE) and transverse magnetic (TM) guided modes. First, we find the eigenmode solutions and obtain the effective refractive indices at 1550 nm wavelength of 500 nm-thick AlN thin film using the Lumerical Mode solution tool. Fig. 1(b) shows the simulation result of the effective indices of the four lowest order modes ($TE_{00}$, $TM_{00}$, $TE_{01}$ and $TM_{01}$). When the waveguide width increases, the electromagnetic field is more strongly confined inside of AlN than the surrounding $SiO_2$ cladding, which results in the increased effective index and multimode. The waveguide width less than 900 nm suppress the higher-order modes while maintaining the fundamental mode effective refractive indices far away from the $SiO_2$ cladding index of 1.44.

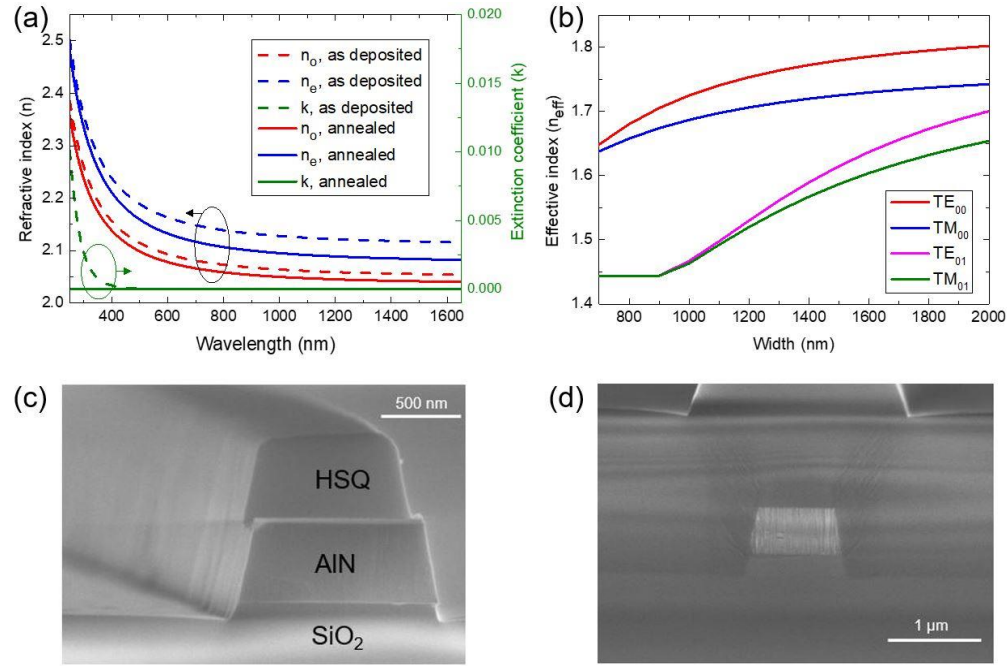

**Fig. 1. (a) Measured complex refractive indices of AlN. (b) The calculated effective indices dependent on the waveguide width. (c) A scanning electron microscope (SEM) image of the AlN waveguide without top cladding, and (d) with $SiO_2$ top cladding.**

The AlN photonic circuits are fabricated as following process. The 500 nm-thick AlN on 3 μm-thick $SiO_2$ on Si substrate is spin-coated with hydrogen silsesquioxane (HSQ) resist (FOX - 15). Then, we expose patterns including straight/curved waveguides and directional couplers using electron-beam lithography (JEOL JBX9300FS). Following that the sample is developed and etched using the inductively coupled plasma reactive ion etching (ICP RIE) with STS Multiplex ICP. The etching gases are $Cl_2$, $BCl_3$ at a rate of 20, 15 sccm, respectively. The etching rate for AlN is about 160 nm/min and 130 nm/min for HSQ at a RF bias power of 100 W and ICP power of 800 W under 5 mTorr pressure. We etch the sample for 4 min which is long enough to fully etch the 500 nm-thick AlN and short enough for 700 nm-thick HSQ to resist.

After the dry etching, AlN waveguide with HSQ e-beam resist is checked with SEM as shown in Fig. 1(c). The AlN is fully etched with enough HSQ on top of the waveguide. To cover the fabricated AlN waveguides, 2 μm plasma enhanced chemical vapor deposition (PECVD) $SiO_2$ is deposited which is enough to protect the optical modes from any optical loss sources on the device. Finally, in order to improve the optical properties of AlN, annealing is performed at 950 °C in $N_2$ environment for one hour. Fig. 1(d) shows the cross section taken by SEM of the final AlN waveguide device with PECVD $SiO_2$ upper cladding and thermal $SiO_2$ under cladding. The device structures in this paper are the same with Fig. 1(d) if not otherwise mentioned.

Normally, the fabricated optical waveguides are not perfect rectangular but have some sidewall angle mainly due to the dry etching process. Especially for the coupling waveguides in which two waveguides are close each other, the sidewall angle is different from the single waveguide's one owing to the etch-lag effect [22]. Considering these fabrication errors, coupling waveguide widths and gaps are carefully designed.

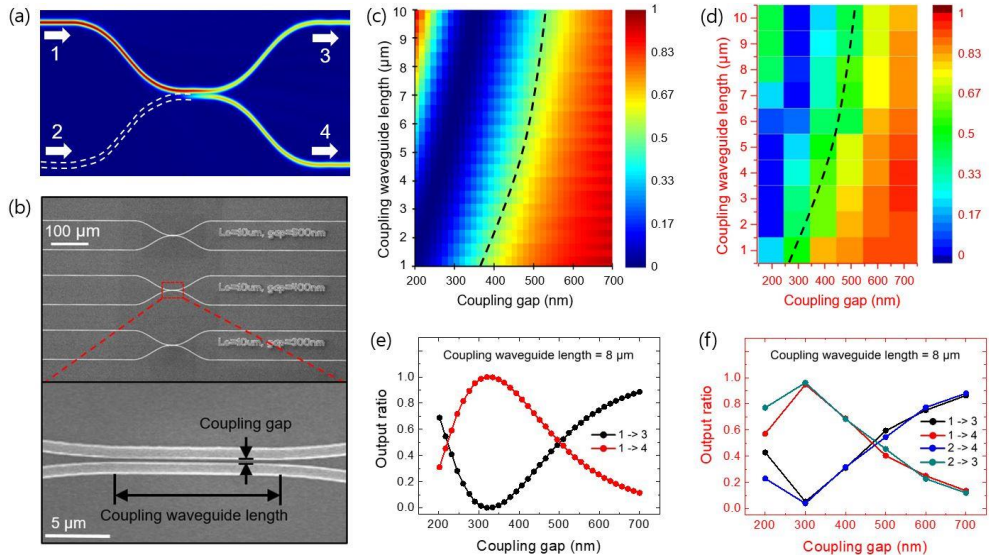

**Fig. 2.** Simulation, fabrication and measurement results of directional couplers at 1550 nm wavelength and TE mode. (a) Directional coupler simulation showing 50:50 output ratio. (b) SEM image of fabricated directional couplers. (c) Simulation and (d) experiment results of the through port (1→3) output ratio map obtained by changing the coupling waveguide length and gap. The black dashed lines indicate a coupling condition of 50:50 ratio. (e) Simulation and (f) experiment results of the output ratios at through (1→3) and coupling (1→4) ports depend on coupling gap when a fixed coupling waveguide length of 8 μm.

## 3. Beam splitter

Beam splitters are one of the key components for photon based information processing system including classical and quantum experiments. Optical waveguide beam splitters generally use evanescent coupling between two adjacent waveguides, so that some portion of the propagating light from one optical waveguide transfer to the other optical waveguide. These beam splitters or directional couplers can also be utilized as frequency filters [23-24], mode converters [25-27], and polarization beam splitters [28-30] by additional waveguide engineering.

The directional coupler in Fig. 2(a) shows when the input power splits to 50:50 at the output ports, which depends on the coupling gap and coupling waveguide length. Fig. 2(b) shows top-view of the SEM images of the fabricated AlN waveguide directional couplers. To study the output beam splitting ratio of AlN waveguide, various directional couplers with the coupling gap from 200 nm to 700 nm and the coupling waveguide length from 1 μm to 11 μm are fabricated while the waveguide width and height are fixed to 800 nm and 500 nm, respectively.

The simulation result of output beam splitting ratio after the directional coupler is shown in Fig. 2(c). The coupling gap and length are scanned while the through port power ratio is monitored. If the coupling gap is too large and the coupling waveguide length is too small, the evanescent wave coupling is too weak to split the input light. Therefore, the two waveguides cannot interact with each other, and 100 % output power at through port is observed at bottom-right area in the Fig. 2(c). If the coupling gap is too small and the coupling waveguide length is too long, the coupling strength becomes too strong. Thus, the coupled light couples back to its original waveguide, so that the optical power between the two adjacent waveguides is periodically exchanged along with the propagation as shown in Fig. 2(c) top-left area and Fig. 2(e). The green area indicate 50:50 output coupling ratio which is the most useful for many optics experiments.

The experiment data with different coupling gap and length are summarized in Fig. 2(d). All fabricated directional couplers are tested using a 1550 nm wavelength laser with TE polarization, and two lensed fibers are used for fiber-to-chip input and output coupling. We observe various output ratios from 50:50 to 1:99 with different coupling gaps and lengths, which matches very well with the simulation result in Fig. 2(c). Fig. 2(e) shows the simulation result of the through and coupled port power ratio at a specific coupling waveguide length of 8 μm. The coupling gap dramatically change the output ratio. Fig. 2(f) shows the measurement result of all four possible coupling cases at a coupling waveguide length of 8 μm. The simulation and experiment results match well except when the coupling gap is 200 nm, which is due to low fabrication tolerances. Based on this reference, we can realize the directional coupler with a desired splitting ratio for the beam splitter function in AlN integrated photonics.

## 4. Polarization beam splitter

We use TE mode for the above experiment, but TM mode has different beam splitting ratio owing to the different effective index. By utilizing this, polarization beam splitters (PBS) are realized via long coupling waveguides and the different coupling strengths between TE and TM modes.

The mode in coupling waveguides can be interpreted as the overlaps of even and odd super-modes [31]. During the two super-modes co-propagation, the light in one waveguide is completely transferred to another waveguide when the phase difference between the even and odd super-modes is π. Therefore, the transfer distance ($D_t$), which is the minimum coupling wavelength length for completely transferring light to the adjacent waveguide, can be expressed using the effective index difference ($\Delta n_{eff} = n_{eff}^{even} - n_{eff}^{odd}$) between the even and odd super-modes for each polarization as below:

$$D_t^{TE} = \frac{\lambda}{2\Delta n_{eff}^{TE}} \qquad (1)$$

$$D_t^{TM} = \frac{\lambda}{2\Delta n_{eff}^{TM}} \qquad (2)$$

, where $\lambda$ is the light wavelength in vacuum, $D_t^{TE}$ and $D_t^{TM}$ are the transfer distances of TE and TM polarizations, respectively. If the coupling waveguides length matches with the integer of the transfer distance ($D_t$), the light transfer to the adjacent waveguide occurs periodically. As the TE and TM modes have different effective indices, their transfer distances are different as well. Thus, we can find the length of coupling waveguide which splits the TE and TM modes to different output ports. In our design, $D_t^{TE}$ is larger than $D_t^{TM}$, and the transfer distance for PBS ($D_{PBS}$) can be expressed as follow:

$$D_{PBS} = nD_t^{TE} = (n+m)D_t^{TM} \quad for\ n = 1, 2, 3\ldots, m = 1, 3, 5\ \ldots \quad (3)$$

To minimize the device footprint, we fix the *m* to 1, which is also a practical solution. Then, we define the polarization diversity parameter $\Delta D_t$ as follow:

$$\Delta D_t = \frac{nD_t^{TE} - (n+1)D_t^{TM}}{D_t^{TM}} \quad for\ n = 1, 2, 3, \ldots \quad (4)$$

The normalized unitless parameter $\Delta D_t$ needs to be 0 to satisfy the PBS. Fig. 3(a) shows the simulation result of $\Delta D_t$ depending on the coupling gap for different *n* at 1550 nm wavelength. There are specific gaps that satisfy $\Delta D_t = 0$ for different *n*. For example, when *n* = 4 (magenta line, $4D_t^{TE} = 5D_t^{TM}$), the line cross 0 when a coupling gap is near 600 nm, which means the TE mode couples back and forth four times to the adjacent waveguide while the TM mode couples five times with the same coupling waveguide length.

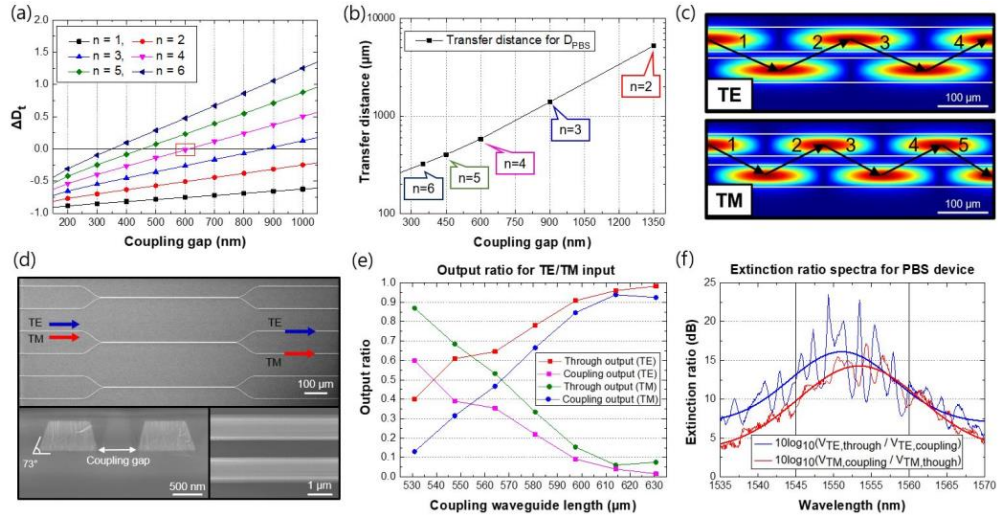

Fig. 3. (a) Calculated **ΔD_t** dependent on the coupling gap at 1550 nm wavelength. (b) Calculated PBS transfer distance (**D_PBS**) with different *n*, thus different coupling gap. (c) The coupling simulation in the case of *n*=4 condition when TE or TM modes are lauched in the upper waveguide. Waveguide width: 700 nm. (d) SEM images of fabricated directional couplers for PBS. (up) top view, (bottom left) cross section of coupling part, (bottom right) top view zoom in of coupling part. (e) Measured output ratio for TE/TM input with different coupling waveguide lengths at 1550 nm wavelength and 600 nm gap. (f) Measured

extinction ratios and their gaussian fitting for TE (blue) and TM (red) inputs from the PBS with a gap of 600 nm and a coupling waveguide length of 614 µm.

The smaller gaps require the larger integer $n$ to satisfy $D_{PBS}$, owing to the small effective index difference of TE and TM modes. However, the small gap enhances the coupling strength so that the overall transfer distance for PBS is shortened as shown in Fig. 3(b). Considering the fabrication tolerance and footprint, $n = 4$ condition is selected which requires a gap of ~ 600 nm and a coupling waveguide length of ~ 600 µm. Fig. 3(c) shows the TE and TM modes coupling simulation from the above parameters. While the TE mode cross the waveguide four times, the TM mode cross the waveguide five times with the same transfer distance, which results in polarization dependent beam splitting. From the simulated parameters, we fabricate AlN PBS based on directional coupler as shown in Fig. 3(d). The SEM images show the top view (up), cross section (bottom left) and zoom-in (bottom right) of the couplers. The fabricated waveguide width at half-height is 850 nm and the sidewall angle is about 73 °.

The measurement results are shown in Fig. 3(e) and (f). The Fig. 3(e) shows the output ratio of TE and TM modes with various coupling waveguide lengths and a fixed coupling gap of 600 nm. The red and magenta squares indicate the TE mode output power ratios at the through and coupled ports, respectively, and the green and blue circles indicate the TM mode output power ratios at the through and coupled ports, respectively. When the coupling waveguide length is 614 ~ 630 µm, the TE mode propagates to the through port while the TM mode goes to the coupled port with extinction ratio > 10 dB at 1550 nm wavelength.

As the coupling is sensitive to the wavelength, we test the wavelength tolerance using the device with a gap of 600 nm and a coupling waveguide length of 614 µm, as shown in Fig. 3(f). The PBS shows high extinction ratio (> 10dB) within 1545 – 1560 nm wavelength range for both TE and TM modes. We expect higher extinction ratio with better wavelength tolerance when the cascaded PBS is realized.

## 5. Quantum experiment

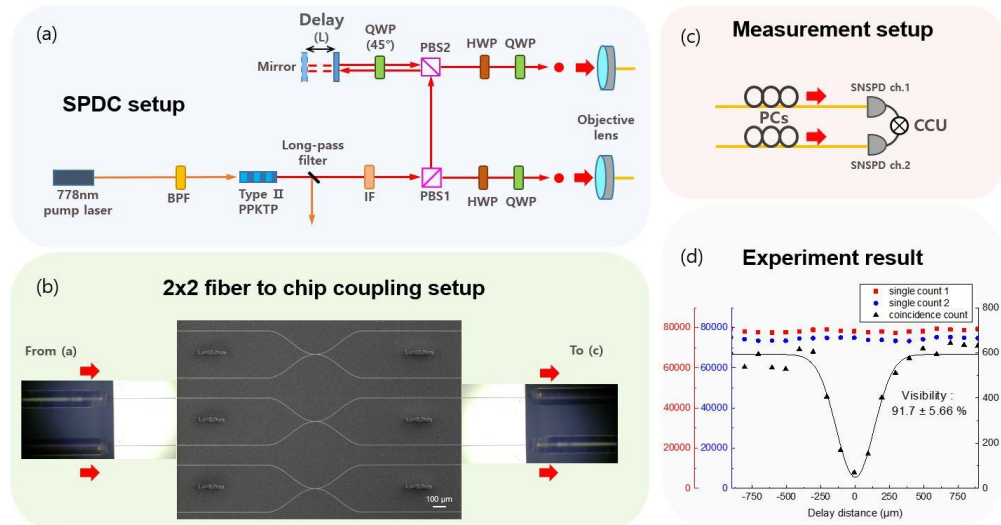

Fig. 4. (a) SPDC setup for single photon pairs generation. (b) Lensed fibers to chip coupling setup scheme of two inputs and two outputs. (c) Two-fold coincidence counting measurement setup. (d) HOM interference result from the AlN waveguide beam splitter. The integration time of photon count for y-axis is 10 sec.

Now, using the fabricated AlN directional coupler with an output ratio of 50:50 in TM mode, we measure two-photon quantum interference using a single photon pair generated from a spontaneous down-conversion (SPDC) and observe the HOM interference [32].

Figure 4 (a) shows the SPDC setup for identical two photons generation at 1556 nm wavelength by 10 mm-long type-II periodically-poled KTP crystal pumped by femtosecond laser pulses with 125 MHz repetition rate. The generated two orthogonally polarized photons (single and idler photons) pass through an interference filter (IF) with 3 nm bandwidth and split by a free space polarizing beam splitter (PBS) 1. The input delay ($L$) for the idler photon is adjusted using a translational stage after PBS2 and quarter-waveplate (QWP) at 45 °. Additional quarter- and half-waveplates (HWP) before the fiber channels ensure the TM polarization at the AlN directional coupler in Fig.4 (b).

Figure 4 (b) shows the SEM image of AlN beam splitters and the picture of the actual 2x2 fiber to chip coupling setup. Four independently adjustable lensed fibers are used for fiber-to-chip coupling. The estimated coupling loss is about 3.7 dB/facet, and the total propagation loss is ~2 dB. The single photon pairs come from the left two optical fibers and go out to the right two optical fibers which is connected to the measurement set up in Fig. 4 (c).

We use superconducting nanowire single photon detectors (SNSPD) which have about 80 % single-photon detection efficiency. The polarization state is optimized by polarization controllers (PCs) in front of the SNSPD. The two-fold coincidence counts are registered by a home-made coincidence counting unit (CCU) [33]. When the optical delay difference between the two paths are zero, two photons come out from the same outputs together, either upper or lower ports. By this HOM interference, there would be no coincidence counts between the two SNSPD channels. The measured HOM interference visibility from the actual counts is 91.7 ± 5.66 % without any post-processing such as accidental coincidence counts subtraction (Fig. 4 (d)). Our experimental results achieve the interference visibility beyond the classical limits, (visibility ≥ 50 %) [34-35], which show that quantum optical experiments can be performed with AlN directional couplers.

## 6. Conclusion

In conclusion, we fabricate a set of AlN optical waveguide directional couplers which covers a variety of splitting ratios from 50:50 to 1:99 by engineering the coupling gap and length. Utilizing the splitting ratio difference of the TE and TM modes in directional couplers, AlN waveguide polarization beam splitters are also realized by simply increasing the coupling waveguide length. Furthermore, a two-photon interference – the Hong-Ou-Mandel (HOM) effect – is demonstrated using the 50:50 directional coupler that proves the potential of AlN photonics platform for quantum optical applications. This study can be a reference for the design and fabrication of AlN devices including directional couplers which is a critical component especially for quantum photonics. The addition of fast modulation using the Pockels effect of AlN would enhance the functionality of the AlN devices for quantum applications, and we expect the quantum information processing devices such as quantum computing and quantum key distribution devices would be realized using the AlN platform in the near future.

**Acknowledgements**

This work was supported by the National Research Foundation of Korea (NRF) (2019M3E4A1079777, 2021M1A2A2043892), the Institute for Information and Communications Technology Promotion (IITP) (2020-0-00947, 2020-0-00972), National Research Council of Science and Technology (NST) (CAP21031-200), and the KIST research program (2E31531).

**Disclosures**

The authors declare no conflicts of interest.